\documentstyle[12pt]{article}

\begin{document}

\baselineskip=24pt 
\renewcommand{\thefootnote}{\fnsymbol{footnote}} 

\begin{titlepage}

\begin{flushright}
FSU-HEP-930924-R \\
Revised January 1994
\end{flushright}

\vspace{0.24in}

\begin{center}

{\bf PRODUCTION OF CP-ODD HIGGS BOSONS \\
WITH LARGE TRANSVERSE MOMENTUM \\
AT HADRON SUPERCOLLIDERS}

\vspace{0.24in}

Chung Kao\footnote{Internet Address: Kao@Fsuhep.physics.fsu.edu}

{\sl
Department of Physics, B-159, Florida State University \\
Tallahassee, FL 32306-3016, USA }

\end{center}

\vspace{0.36in}

\begin{abstract}

A two Higgs doublet model is employed to study the production
of a CP-odd Higgs boson ($A$) associated
with a large transverse momentum jet ($j$) at hadron supercolliders.
The cross section of $pp \to jA+X$ is evaluated
with four subprocesses:
$gg \to gA$, $gq \to qA$, $g\bar{q} \to \bar{q}A$ and $q\bar{q} \to gA$.
We find that $pp \to jA+X$ is a significant source of
CP-odd Higgs bosons at future hadron supercolliders.

\end{abstract}

\end{titlepage}


\noindent{\bf 1. Introduction}

In the Standard Model (SM) of electroweak interactions,
only one Higgs doublet is required to generate masses for the fermions
as well as the gauge bosons. A single neutral CP-even Higgs boson ($H^0$)
appears after the spontaneous symmetry breaking.
Various extensions of the SM have more Higgs multiplets
and lead to additional physical spin-0 fields \cite{Guide}.

A general two Higgs doublet model \cite{Georgi} has doublets $\Phi_1$
and $\Phi_2$ with vacuum expectation values (VEV's) $v_1$ and $v_2$.
If CP is invariant in the Higgs sector,
there remain five physical Higgs bosons \cite{Guide} after symmetry breaking:
a pair of singly charged Higgs bosons $H^{\pm}$, two neutral CP-even scalars
$H$ (heavier) and $h$ (lighter), and a neutral CP-odd pseudoscalar $A$.

Two models with a discrete symmetry \cite{GW} have been considered
for the Yukawa interactions among the Higgs bosons and fermions.
In model I \cite{Model1,Model12}, all fermions couple to $\Phi_2$,
and the $Aq\bar{q}$ interaction takes the form
\begin{equation}
{\cal L}_{Aq\bar{q}} = +i\frac{m_u}{v} \cot \beta \bar{u} \gamma_5 u A
                       -i\frac{m_d}{v} \cot \beta \bar{d} \gamma_5 d A
\end{equation}
In model II \cite{Model2,Model12}, which is required in the minimal
supersymmetry model (MSSM)\footnote{Reviews of the MSSM can be found in
\cite{Nilles}-\cite{Tata}.
Recent studies on the search for MSSM Higgs bosons at hadron supercolliders
are to be found in \cite{Barger}-\cite{Dai}. },
$\Phi_1$ couples to down-type quarks
and charged leptons while $\Phi_2$ couples to up-type quarks and neutrinos,
and the $Aq\bar{q}$ interaction is
\begin{equation}
{\cal L}_{Aq\bar{q}} = +i\frac{m_u}{v} \cot \beta \bar{u} \gamma_5 u A
                       +i\frac{m_d}{v} \tan \beta \bar{d} \gamma_5 d A
\end{equation}
where $\tan \beta \equiv v_2/v_1$, $\sqrt{ v_1^2 +v_2^2 } = v = 2M_W/g$,
$u$ and $d$ are generic u-type and d-type quarks.
Without loss of generality, Model II is chosen in all our calculations.
In this model, the $Ad\bar{d}$ coupling is enhanced when $\tan \beta$ is large.
The cross section for Model I is usually dominated by the contribution from
the top quark, and can be well approximated by our data at $\tan \beta = 1$
multiplied by $\cot^2 \beta$.

At hadron supercolliders, such as the SSC in the USA and the LHC at CERN,
the SM Higgs boson can be produced dominantly from gluon fusion \cite{gluon},
and from vector boson fusion \cite{Cahn}-\cite{Kane}.
The $A$ does not couple to the gauge bosons at the tree level,
therefore, gluon fusion and heavy quark fusion \cite{Duane}
are the two major sources of the $A$ in high energy hadron collisions.

\medskip

\noindent{\bf 2. CP-odd Higgs Bosons with Large $P_T$}

If the SM Higgs boson ($H^0$) can be produced in association
with a large transverse momentum jet ($j$), $pp \to jH^0+X$,
via the following subprocesses: $gg \rightarrow gH^0$, $gq \rightarrow qH^0$,
$g\bar{q} \rightarrow \bar{q}H^0$, and $q\bar{q} \rightarrow gH^0$,
the mass of the Higgs boson might be reconstructed from its
$\tau^+\tau^-$ decay channel \cite{Ellis} if the Higgs mass is in the
intermediate range; or from the $ZZ$ decay mode \cite{Baur}\footnote{
I would like to thank Uli Baur for comparing the matrix elements
in this reference. }
if the Higgs boson is heavier.

The same subprocesses can also produce the CP-odd Higgs bosons
with large transverse momentum ($P_T$) via triangle and box diagrams,
as shown in Figure 1.
The amplitudes of the quark loops can be expressed
in terms of form factors \cite{Tini} which are combinations of
scalar one-loop integrals \cite{HV}.
We have calculated all loop integrations with the computer
code LOOP \cite{LOOP}, which evaluates one loop integrals
analytically and generates numerical data.

If there exist just three generations of quarks,
only the top quark loop ($t$-loop) and the bottom quark loop ($b$-loop)
make significant contributions.
Therefore, only the third generation quark loops are considered
in our calculations with the following values of parameters:
$\alpha = 1/128$, $M_Z = 91.17$ GeV, $M_W = 80.0$ GeV,
$\sin^2 \theta_W = 0.230$, the bottom quark mass $m_b =$ 5 GeV.
If not specified, the top quark mass ($m_t$) is taken to be 150 GeV.
The mass of the CP-odd Higgs boson ($m_A$) is considered to be between
50 and 1000 GeV.
The updated parton distribution functions of Owens \cite{Owens}
with $\Lambda = 0.177$ GeV and $Q^2 = m_A^2 +P_T^2$ are chosen
to evaluate the cross section of $pp \to jA+X$  at the energies
of the SSC ($\sqrt{s} = 40$ TeV) and the LHC ($\sqrt{s} = 14$ TeV).
Since it is at the order of $\alpha_s^3$, the cross section
of $pp \to jA+X$ is very sensitive to the choice of $\Lambda$ and $Q^2$.
To evaluate the production rate of $A$ with large $P_T$
as well as to avoid the singularities at $P_T \to 0$, we impose
a $P_T$ cut on the $A$ and the jet: $P_T > 50$ GeV.

Figure 2 shows the cross section of $pp \to jA+X$
at the SSC and the LHC energies, as a function of
$m_A$, for $m_t = 150$ GeV and various values of $\tan \beta$.
The cross section is about 5-19 times larger
at $\sqrt{s} = 40$ TeV than at $\sqrt{s} = 14$ TeV,
for 50 GeV $< m_A < 1000$ GeV.
The loop integrals are functions of the mass of the CP-odd Higgs boson ($m_A$),
the quark mass in the loops ($m_q$),
and the Mandelstam variables: $\hat{s}, \hat{t}, \hat{u}$.
For $m_A < m_t$, the $t$-loop is almost independent of $m_t$,
thus the amplitude is very sensitive to $\tan \beta$.
For $\tan \beta$ larger than about 5, the cross section is dominated by
the $b$-loop.
At the threshold of $m_A = 2m_t$, the imaginary part of the amplitude
is turned on. Therefore, the amplitude squared ($|M|^2$)
grows rapidly when $m_A$ is close to $2m_t$.
When $m_q^2 \ll m_A^2 < \hat{s}$, the amplitude squared behaves as
\begin{equation}
|M_{q-loop}|^2 \sim m_q^4 ln^4(m_q^2).
\end{equation}
The $t$-loop dominates in a large region of $\tan \beta$.
The cross section is almost proportional to $\cot^2\beta$ for $\tan \beta <
10$.
Only for $\tan\beta$ close to $m_t/m_b$, can the $b$-loop dominate and the
total cross section be enhanced by large $\tan \beta$.
Not shown is the interference between the $t$-loop and the $b$-loop.
The $t$-loop and the $b$-loop interfere destructively if $m_A$ is close to
$2m_t$, but constructively if $m_A$ is away from $2m_t$.

To compare the production rate of the CP-odd Higgs boson ($A$)
to that of the SM Higgs boson ($H^0$) at the SSC energy,
we present the cross section of $pp \to \phi+X$, $\phi=H^0$ or $A$
from various subprocesses in Table I, for $m_t = 150$ GeV and $\tan \beta =1$.
Several interesting aspects can be learned from Table I and Figure 2:
(1) If the Higgs bosons are produced from $gg \to \phi$, or the subprocesses
of $pp \to j\phi+X$ via quark loops,
the cross section of $A$ is at least twice that of the $H^0$
for $m_\phi < 500$ GeV.
At $m_\phi = 2m_t$,
the cross section of $A$ is about 5 times that of the $H^0$;
which implies that the threshold enhancement at $2m_t$ is much larger
for the $A$ than for the $H^0$.
For larger $m_\phi$, their cross sections are about the same.
(2) $gg \to g\phi$ dominates and contributes about $80\%$
to the cross section of $pp \to j\phi+X$.
(3) For $m_\phi > 400$ GeV, the number of Higgs bosons produced
from $gg \to g\phi+X$ with $P_T(\phi) > 50$ GeV, is almost comparable
to that from $gg \to \phi$.
(4) For $m_\phi > 50$ GeV, the cross sections of the $A$ and $H^0$ are
the same from $gg \to \phi b\bar{b}$, which is a good approximation
to the `exact' cross section of $\phi$ produced from $b\bar{b}$ fusion
\cite{Duane}.
(5) For $\tan\beta > 10$, $gg \to Ab\bar{b}$ becomes the major source
of large $P_T$ CP-odd Higgs bosons.
Its cross section is proportional to $\tan^2\beta$ and it is greatly enhanced
by large $\tan\beta$.

The effects of $\tan \beta$ and $m_t$ on the cross section
of $pp \to jA+X$ at $\sqrt{s} = 40$ TeV are shown in Figure 3,
for $m_t = 120, 150$ and 180 GeV.
Two values of $m_A$ are considered:
(a) $m_A = 200$ GeV, which is less than $2m_t$; and
(b) $m_A = 400$ GeV, which is larger than $2m_t$.
If $\tan \beta$ is less than about 10,
$m_t$ slightly affects the total cross section:
larger $m_t$ slightly enhances the cross section for $m_A = 200$ GeV,
but slightly reduces the cross section for $m_A = 400$ GeV.
The total cross section is almost independent of $m_t$ for $\tan\beta > 10$.

The top quark mass dependence on the matrix element squared of $gg \to gH^0$
has been studied in detail \cite{Ellis,Baur}. A similar study for $gg \to gA$
is currently under investigation. At the threshold of $2m_t$, the enhancement
on the $|M|^2$ for $gg \to g\phi$ is very similar to that of $gg \to \phi$,
where $\phi = H^0$ or $A$. Therefore, we review and discuss the production of
$pp \to \phi+X$ from gluon fusion in the appendix.

In Figure 4, we show the $P_T$ distribution of $pp \to gA+X$ from $gg \to gA$
at $\sqrt{s} = 40$ GeV for $m_t = 150$ GeV, $m_A = 200$ and 400 GeV.
The effect of $\tan \beta$ is similar to that on the cross section
versus $m_A$ in Fig. 2.
If we require $P_T(A) > 100$ GeV,
the cross section will be reduced to only about
1/3 of that with $P_T > 50$ GeV for $m_A < 200$ GeV
while about 1/2 of $m_A > 400$ GeV will survive.

\medskip

\noindent{\bf 3. Large Quark Mass Limit}\footnote{
In this section, $\tan \beta$ is taken to be 1;
$\alpha_s$ is taken to be $\frac{12\pi}{23}/ln( Q^2/\Lambda^2)$;
and only the top quark loop is considered.}

If the quark mass in the loop diagrams is much larger than that of
the CP-odd Higgs boson, $m_q \gg m_A$,
the $ggA$ and $gggA$ couplings can be obtained from the low energy
theorem of the axial anomaly \cite{Adler}-\cite{Bill} or
from the exact calculation of $gg \to A$ at the limit of $m_q^2/m_A^2 \gg 1$.
An effective Lagrangian \cite{Zerwas1,Kauffman} can be written as
\begin{eqnarray}
{\cal L}_{eff} = \frac{\lambda}{8} G^a_{\mu\nu} \tilde{G}^{a\mu\nu} A
\end{eqnarray}
where $\lambda = \frac{\alpha_s}{2 \pi v}$ and
$\tilde{G}^{a\mu\nu} = \epsilon^{\mu\nu\rho\sigma} G^a_{\rho\sigma}$.

Applying the effective Lagrangian, we obtain the amplitude squared
for various subprocesses at the large quark mass limit
\begin{eqnarray}
|M(gg \to gA)|^2       & = & \frac{3f}{32} (\frac{s^4+t^4+u^4+m_A^8}{stu}) \\
|M(gq \to qA)|^2       & = & \frac{f}{24} [-(\frac{s^2+u^2}{t})] \\
|M(q\bar{q} \to gA)|^2 & = & \frac{f}{9} (\frac{t^2+u^2}{s})
\end{eqnarray}
where $f = \lambda^2g_s^2$ and the $|M|^2$ has been summed and averaged
over all spins and colors.

Table II shows the $P_T$ distribution ($d\sigma/dP_T$) of $pp \to gA+X$,
at $\sqrt{s} = 14$ TeV, from the top quark loops of
$gg \to gA$, for various $m_t$ and $m_A$, with $\tan \beta = 1$.
The large quark mass limit is a very good approximation for $P_T < m_t$ and
$m_A < m_t$. It provides a good check for the exact calculation.
It slightly underestimates the exact cross section if $m_A$ is close to $2m_t$
and $P_T$ is less than $2m_t$, because the exact cross section is enhanced by
the threshold effect.
However, it overestimates the cross section if $P_T$ is larger than $2m_t$
or if $m_A$ is larger than about $4m_t$. Similar results have been
found for the SM Higgs boson \cite{Ellis,Baur}.

\medskip

\noindent{\bf 4. Conclusions}

The $pp \to jA+X$ is a very significant source of CP-odd Higgs bosons
at future hadron colliders for $\tan \beta$ less than about 10,
especially for the detection modes that require a large $P_T$ for the $A$.
The large quark mass limit of $pp \to jA+X$ is a good approximation
for $m_A < m_t$, but overestimates the exact cross section for
large $m_A$ and $P_T$.
In the region of small $P_T$, a complete description requires
higher order corrections \cite{Kauffman}-\cite{Zerwas2},
and the resummation of gluon emission \cite{resummation}-\cite{Yuan}.

If the $Ab\bar{b}$ coupling is proportional to $\tan\beta$,
$gg \to Ab\bar{b}$ is the dominant process to produce the $A$ for
$\tan \beta > 10$ and $m_A > 100$ GeV.
The subprocesses $gg \to gA$ and $gg \to Ab\bar{b}$ can be considered
as complementary to each other for producing large $P_T$ CP-odd Higgs bosons
at future hadron colliders.

The total cross section presented in this letter for $pp \to jA+X$
is less reliable for a much heavier $A$.
It is likely to overestimate the production rates due to
large $ln(m_A^2/P_{T0}^2)$ contributions,
since a constant minimal transverse momentum cutoff ($P_T > P_{T0} = 50$ GeV)
has been applied.

The total cross section of $pp \to jA+X$ at $\sqrt{s} = 14$ TeV
is about 5-19 times smaller than at $\sqrt{s} = 40$ TeV,
for 50 GeV $< m_A < 1000$ GeV.

\medskip

\noindent{\bf Acknowledgements}

I am grateful to Howie Baer, Bill Bardeen, Uli Baur, Joe Polchinski
and Xerxes Tata for beneficial discussions, to Sally Dawson and
Duane Dicus for continuing encouragement as well as valuable comments
and to Harvey Goldman for technical support.
This research was supported in part by the DOE contract DE-FG05-87-ER40319.

\newpage

\noindent{\bf Appendix}\footnote{
In this appendix, $m = m_t$; $\tan \beta = 1$; and
$\phi$ is the SM Higgs boson $H^0$ or a CP odd Higgs boson $A$.}

At the lowest order, the cross sections of $gg \to \phi$
via the top quark loops are
\begin{eqnarray}
\sigma(gg \to H^0)& = & \frac{1}{64} (\frac{\alpha_s^2 \alpha_W}{M_W^2})
                                     (s) |F(\rho)|^2 \delta(s-{M_H^0}^2)
                             \nonumber \\
\sigma(gg \to A)  & = & \frac{1}{64} (\frac{\alpha_s^2 \alpha_W}{M_W^2})
                                         (s) |G(\rho)|^2 \delta(s-M_A^2)
                                     \nonumber
\end{eqnarray}
where $\alpha_W = \alpha/\sin^2 \theta_W$ and $\rho = m^2/m_\phi^2$.

The functions $F(\rho)$ and $G(\rho)$ are
\begin{eqnarray}
F(\rho)   & = & +\rho[ 2 + (4\rho -1)I(\rho)] \nonumber \\
G(\rho)   & = & -\rho I(\rho) \nonumber
\end{eqnarray}
and the function $I(\rho)$ is
\begin{eqnarray}
I(\rho) & = & +\int_{0}^{1} \frac{dy}{y}\{ln[1-\frac{y(1-y)}{\rho-i\epsilon}]\}
               \nonumber \\
        & = & -2 [\sin^{-1}(\frac{1}{2\sqrt{\rho}})]^2, \rho \geq \frac{1}{4}
               \nonumber \\
        &   & +\frac{1}{2} [ln(\frac{z_+}{z_-})-i\pi]^2, \rho < \frac{1}{4}
               \nonumber
\end{eqnarray}
where $z_{\pm} = [1 \pm \sqrt{1-4\rho} ]/2$.

I. When $m$ is very small, $\rho \ll 1$, therefore,
\begin{eqnarray}
I(\rho) =  \frac{1}{2} [ln^2(\frac{s}{m^2}) -\pi^2
                             -2i\pi ln(\frac{s}{m^2}) ]. \nonumber
\end{eqnarray}
II. At the threshold, $\rho = 1/4$, therefore
\begin{eqnarray}
I(\frac{1}{4}) & = & -\pi^2/2, \nonumber \\
F(\frac{1}{4}) & = & +\frac{1}{2}, \nonumber \\
G(\frac{1}{4}) & = & +\frac{\pi^2}{8}. \nonumber
\end{eqnarray}
III. At very large $m$, $\rho \gg 1$, therefore,
\begin{eqnarray}
I(\rho) & = & -\frac{1}{2\rho} -\frac{1}{24\rho^2} +O(\frac{1}{\rho^3}),
                \nonumber \\
F(\rho) & = & +\frac{1}{3} +O(\frac{1}{\rho}), \nonumber \\
G(\rho) & = & +\frac{1}{2} +O(\frac{1}{\rho}). \nonumber
\end{eqnarray}

Therefore, at the threshold of $m_\phi = 2m_t$,
\begin{eqnarray}
\frac{\sigma(gg \to A)}{\sigma(gg \to H^0)} = \pi^4/16 \simeq 6; \nonumber
\end{eqnarray}
while at the large top quark mass limit,
\begin{eqnarray}
\frac{\sigma(gg \to A)}{\sigma(gg \to H^0)} = 9/4. \nonumber
\end{eqnarray}
The same ratios appear in the cross sections of $gg \to g\phi$
at both limits.

\newpage
%

\newpage
\noindent{\bf Tables }

\bigskip

TABLE I.
The cross section of $pp \to \phi+X$ at $\sqrt{s} = 40$ TeV, in $pb$,
as a function of $m_\phi$, where $\phi$ is the SM Higgs boson ($H^0$)
or the CP-odd Higgs boson ($A$).
Various subprocesses are considered with $m_t = 150$ GeV and $\tan\beta = 1$.
We have imposed a cut on $P_T (\phi) = P_T > 50$ GeV for $pp \to j\phi+X$
and $pp \to \phi b\bar{b}+X$.

\medskip

\begin{center}
\begin{tabular}{lccccccc}
\hline
$m_\phi$(GeV)   & 50 & 100 & 200 & 300 & 400 & 600 & 800 \\
\hline
Raw Cross Section \\
$gg \to H^0$              & 310 & 120 & 52 & 47 & 40 & 9.6 & 2.7 \\
$gg \to H^0t\bar{t}$ & 27 & 7.7 & 1.4 & 0.68 & 0.44 & 0.21 & 0.11 \\
$gg \to H^0b\bar{b}$ & 43 & 7.8 & 1.0 & 0.27 & 0.097 & 0.021
& 6.7$\times10^{-3}$ \\
$P_T >$ 50 GeV \\
$gg \to H^0b\bar{b}$ & 2.5 & 1.0 & 0.26 & 0.091 & 0.039 & 0.010
& 3.6$\times10^{-3}$ \\
$gg \to gH^0$             & 64 & 40 & 23 & 22 & 21 & 6.2 & 2.0 \\
$gq \to qH^0$             & 10 & 6.5 & 3.7 & 3.6 & 3.3 & 1.0 & 0.33 \\
$g\bar{q} \to \bar{q}H^0$ & 4.4 & 2.7 & 1.4 & 1.3 & 1.1 & 0.31 & 0.092 \\
$q\bar{q} \to gH^0$       & 0.17 & 0.12 & 0.041 & 7.2$\times10^{-3}$
& 1.9$\times10^{-3}$ & 2.8$\times10^{-4}$ & 6.7$\times10^{-5}$ \\
\hline
Raw Cross Section \\
$gg \to A$  & 620  & 270 & 130 & 290 & 87 & 14  & 3.6 \\
$gg \to At\bar{t}$ & 6.9 & 4.6 &  2.2 & 1.2 & 0.71 & 0.29 & 0.13 \\
$gg \to Ab\bar{b}$ & 46 & 7.9 & 1.0 & 0.27 & 0.097 & 0.021
& 6.7$\times10^{-3}$ \\
$P_T >$ 50 GeV \\
$gg \to Ab\bar{b}$ & 2.6 & 1.0 & 0.26 & 0.091 & 0.039 & 0.010
& 3.6$\times10^{-3}$ \\
$gg \to gA$ & 136 & 89  & 58  & 125 & 46 & 9.4 & 2.7 \\
$gq \to qA$ & 22  & 15  & 9.6 & 21 & 7.5 & 1.6 & 0.45 \\
$g\bar{q} \to \bar{q}A$ & 9.5 & 6.1 & 3.7 & 7.6 & 2.6 & 0.47 & 0.12 \\
$q\bar{q} \to gA$ & 0.36 & 0.25 & 0.11 & 0.042 & 4.5$\times10^{-3}$
& 5.1$\times10^{-4}$ & 1.1$\times10^{-4}$ \\
\hline
\end{tabular}
\end{center}
%

\newpage

TABLE II.
The $P_T$ distribution ($d\sigma/dP_T$) of $pp \to gA+X$ in pb/GeV,
at $\sqrt{s} = 14$ TeV, as a function of $m_A$ and $m_t$.
We consider only the top quark loop diagrams of $gg \to gA$ with
$\tan \beta = 1$.
\medskip

\begin{center}
\begin{tabular}{lccccccc}
\hline
$m_t \backslash P_T$(GeV)    & 100 & 200 & 400 & 600 & 800 & 1000 \\
\hline
$m_A = 100$ GeV \\
100      & 1.1$\times 10^{-1}$ & 8.9$\times 10^{-3}$ & 2.3$\times 10^{-4}$
         & 1.5$\times 10^{-5}$ & 1.6$\times 10^{-6}$ & 2.4$\times 10^{-7}$ \\
150      & 1.0$\times 10^{-1}$ & 1.2$\times 10^{-2}$ & 4.0$\times 10^{-4}$
         & 3.1$\times 10^{-5}$ & 3.8$\times 10^{-6}$ & 6.0$\times 10^{-7}$ \\
200      & 9.9$\times 10^{-2}$ & 1.3$\times 10^{-2}$ & 5.6$\times 10^{-4}$
         & 4.9$\times 10^{-5}$ & 6.4$\times 10^{-6}$ & 1.1$\times 10^{-6}$ \\
$\infty$ & 9.5$\times 10^{-2}$ & 1.2$\times 10^{-2}$ & 8.1$\times 10^{-4}$
         & 1.2$\times 10^{-4}$ & 2.5$\times 10^{-5}$ & 6.3$\times 10^{-6}$ \\
$m_A = 400$ GeV \\
100      & 1.2$\times 10^{-2}$ & 2.2$\times 10^{-3}$ & 1.4$\times 10^{-4}$
         & 1.4$\times 10^{-5}$ & 1.7$\times 10^{-6}$ & 2.7$\times 10^{-7}$ \\
150      & 3.3$\times 10^{-2}$ & 6.2$\times 10^{-3}$ & 4.4$\times 10^{-4}$
         & 4.3$\times 10^{-5}$ & 5.6$\times 10^{-6}$ & 9.1$\times 10^{-7}$ \\
200      & 6.5$\times 10^{-2}$ & 1.2$\times 10^{-2}$ & 8.5$\times 10^{-4}$
         & 8.8$\times 10^{-5}$ & 1.2$\times 10^{-5}$ & 2.0$\times 10^{-6}$ \\
$\infty$ & 1.2$\times 10^{-2}$ & 2.8$\times 10^{-3}$ & 3.8$\times 10^{-4}$
         & 7.5$\times 10^{-5}$ & 1.8$\times 10^{-5}$ & 5.0$\times 10^{-6}$ \\
$m_A = 600$ GeV \\
100      & 1.8$\times 10^{-3}$ & 3.9$\times 10^{-4}$ & 4.0$\times 10^{-5}$
         & 5.4$\times 10^{-6}$ & 8.5$\times 10^{-7}$ & 1.5$\times 10^{-7}$ \\
150      & 5.0$\times 10^{-3}$ & 1.1$\times 10^{-3}$ & 1.3$\times 10^{-4}$
         & 1.8$\times 10^{-5}$ & 2.9$\times 10^{-6}$ & 5.4$\times 10^{-7}$ \\
200      & 1.0$\times 10^{-2}$ & 2.3$\times 10^{-3}$ & 2.7$\times 10^{-4}$
         & 3.9$\times 10^{-5}$ & 6.5$\times 10^{-6}$ & 1.2$\times 10^{-6}$ \\
$\infty$ & 4.7$\times 10^{-3}$ & 1.2$\times 10^{-3}$ & 2.0$\times 10^{-4}$
         & 4.6$\times 10^{-5}$ & 1.3$\times 10^{-5}$ & 3.8$\times 10^{-6}$ \\
$m_A = 1000$ GeV \\
100      & 1.1$\times 10^{-4}$ & 2.7$\times 10^{-5}$ & 4.0$\times 10^{-6}$
         & 8.1$\times 10^{-7}$ & 1.8$\times 10^{-7}$ & 4.2$\times 10^{-8}$ \\
150      & 3.2$\times 10^{-4}$ & 8.2$\times 10^{-5}$ & 1.3$\times 10^{-5}$
         & 2.7$\times 10^{-6}$ & 6.2$\times 10^{-7}$ & 1.5$\times 10^{-7}$ \\
200      & 6.9$\times 10^{-4}$ & 1.7$\times 10^{-4}$ & 2.9$\times 10^{-5}$
         & 6.2$\times 10^{-6}$ & 1.5$\times 10^{-6}$ & 3.6$\times 10^{-7}$ \\
$\infty$ & 1.1$\times 10^{-3}$ & 3.0$\times 10^{-4}$ & 5.7$\times 10^{-5}$
         & 1.6$\times 10^{-5}$ & 5.2$\times 10^{-6}$ & 1.8$\times 10^{-6}$ \\
\hline
\end{tabular}
\end{center}
%

%
\newpage
\noindent{\bf Figures}

\bigskip

FIG. 1 The Feynman diagrams of the subprocesses
(a) $gg \to gA$, (b) $gq \to qA$ and (c) $q\bar{q} \to gA$.
We have not shown the diagrams with various permutations of the external legs.

\medskip

FIG. 2 The cross section of $pp \to jA+X$ in $pb$ versus $m_A$,
for $m_t = 150$ GeV, $\tan \beta = 1,2,5,10$ and 30,
at the energies: (a) $\sqrt{s} = 40$ TeV and (b) $\sqrt{s} = 14$ TeV.

\medskip

FIG. 3 The cross section of $pp \to jA+X$ in $pb$ versus $\tan\beta$,
at $\sqrt{s} = 40$ TeV, for $m_t = 120$ (dotted), 150(solid)
and 180 (dashed) GeV. Two masses of the CP-odd Higgs boson are considered:
(a) $m_A = 200$ GeV and (b) $m_A = 400$ GeV.

\medskip

FIG. 4 The $P_T$ distribution of $pp \to gA+X$ from
$gg \to gA$ at $\sqrt{s} = 40$ TeV, for $m_t = 150$ GeV
and $\tan\beta = 1,2,5,10$ and 30. Two masses of the CP-odd Higgs boson
are considered: (a) $m_A = 200$ GeV and (b) $m_A = 400$ GeV.

\end{document}